\documentclass[floatfix, reprint,superscriptaddress,amssymb,amsmath,aps,prx,longbibliography]{revtex4-2}
\usepackage{minitoc}

\usepackage{graphicx}
\usepackage{dcolumn}
\usepackage{threeparttable}
\usepackage{bm}
\usepackage{siunitx}
\usepackage{booktabs}
\usepackage[usenames,dvipsnames]{xcolor}
\usepackage{tabularx}
\usepackage{array}
\usepackage{colortbl}
\usepackage{braket}
\usepackage{gensymb}
\usepackage{pdfpages}
\usepackage{amsmath}
\usepackage{mathrsfs}
\usepackage{blindtext}
\usepackage{minitoc}
\usepackage{dsfont}
\usepackage{multirow}
\usepackage{xcolor}
\usepackage{soul}
\usepackage{stfloats}
\usepackage{hyperref}
\usepackage{float}
\usepackage{subfig}

\usepackage{amssymb}
\usepackage{amsthm}

\makeatletter
\patchcmd{\@outputpage@head}{\@ifx{\LS@rot\@undefined}{}{\LS@rot}}{}{}{}
\makeatother

\newcounter{algorithm}

\newcounter{alg}
\renewcommand{\thealg}{\arabic{alg}}

\hypersetup{ 
    pdfauthor={Zhehao Yi},   
    pdftitle={Geometric Optimization on Lie Groups: A Lie-Theoretic Explanation of Barren Plateau Mitigation for Variational Quantum Algorithms},   
    pdfsubject={Quantum Machine Learning}, 
    pdfkeywords={quantum, lie groups},
    pdfcreator={Zhehao Yi}, 
    pdfproducer={Zhehao Yi}, 
     colorlinks=true,
     filecolor=green, 
     citecolor = blue,       
     urlcolor=red, 
     } 

\begin{document}

\title{Geometric Optimization on Lie Groups: A Lie-Theoretic Explanation of Barren Plateau Mitigation for Variational Quantum Algorithms}

\author{Zhehao Yi}
\affiliation{AI, Autonomy, Resilience, Control (AARC) Lab, Electrical \& Computer Engineering, The University of Alabama in Huntsville, AL, 35899, USA}

\author{Rahul Bhadani}
\affiliation{AI, Autonomy, Resilience, Control (AARC) Lab, Electrical \& Computer Engineering, The University of Alabama in Huntsville, AL, 35899, USA}


\begin{abstract}
Barren plateaus, which means the training gradients become extremely small, pose a major challenge in optimizing parameterized quantum circuits, often making the learning process impractically slow or stall. This work shows why using neural networks to generate quantum circuit parameters helps overcome this difficulty. We introduce a geometric viewpoint that describes how the parameters produced by neural networks evolve during training. Our analysis shows that these parameters follow smooth and efficient paths that avoid the flat regions in the training that cause barren plateaus. This provides a computational explanation for the improved trainability observed in recent neural network–assisted quantum learning methods. Overall, our findings bridge ideas from quantum machine learning and computational optimization, offering new insight into the structure of quantum models and guiding future approaches for designing more trainable quantum circuits or parameter initialization.
\end{abstract}

\maketitle

\section{Introduction}
Optimization lies at the core of modern machine learning, and this is equally true in the emerging field of quantum machine learning (QML). In particular, variational quantum algorithms (VQAs) have become the primary framework for training parameterized quantum circuits, where a classical optimizer updates quantum gate parameters to minimize a given cost function. These algorithms are particularly suitable for the noisy intermediate-scale quantum (NISQ) era, offering a hybrid quantum–classical approach for learning and optimization. Applications span from quantum convolutional neural networks for pattern recognition to quantum approximate optimization algorithms (QAOA) for combinatorial optimization, and more recently, quantum transformers for structured data processing \citep{1,2,3,4,5,6,7,8,9,10}.

Despite the advantages of VQAs, it still often suffers from severe optimization pathologies, most notably is the barren plateau \citep{11}. In this situation, gradient magnitudes decay exponentially with the number of qubits, causing the optimization landscape to become nearly flat and rendering gradient-based training ineffective. Consequently, the number of iterations required to converge
to a global optimum grows exponentially, leading to excessive
resource consumption and undermining the potential quantum
advantage of VQAs. 

Several strategies have been proposed to mitigate the barren plateau problem. Among them, a state-efficient ansatz has been shown to improve trainability by modulating the entanglement capacity of the quantum circuit \citep{12}. Additionally, parameter initialization inspired by transfer learning has proven effective in alleviating the barren plateau phenomenon and enhancing the overall trainability of variational quantum algorithms \citep{13}. Furthermore, incorporating Markovian losses into the optimization process can also help improve the training performance of parameterized quantum circuits \citep{14}. Among these approaches, a method using neural networks to generate parameters for parameterized quantum circuits has garnered significant attention \citep{15,16}. This technique involves using neural networks to produce parameters for either fixed or randomized quantum circuits, which has been shown to achieve a fixed cost function value more efficiently or converge faster compared to models initialized with purely random parameters without the use of neural networks. Despite the effectiveness of neural networks in generating quantum gate parameters that can mitigate barren plateaus, the theory for this approach is still unclear. Motivated by this gap, we seek to provide a mathematical analysis of this improved performance.

In this work, we adopt tools from Lie theory to analyze the parameter update process in variational quantum algorithms, providing a geometric interpretation of the optimization in variational quantum algorithms. Specifically, we map the single-qubit quantum gate onto a corresponding $\mathbb S\mathbb U(2)$ Lie group, such that the parameter updates can be interpreted as evolving within the associated $\mathfrak s\mathfrak u(2)$ Lie algebra. From this geometric perspective, the optimization trajectory of the parameters corresponds to one of the geodesics on the Lie group, which allows us to characterize its behavior using the mathematical framework of Lie theory. 
This article is organized as follows. Section \ref{sec: 2} provides the necessary preliminaries on Lie Theory. Section \ref{sec: 3} describes the proposed methodology. Section \ref{sec: 4} presents and analyzes the experimental results, and Section \ref{sec: 5} concludes the paper with a summary and future outlook.

\section{Preliminaries on Lie Theory and $\mathbb{SU}(2)$ group}\label{sec: 2}
In this section, we provide the necessary background material that will be used throughout the subsequent subsections. The primary reference for much of the material in this section is \citep{17,18,ref_alexandrino,bullo1995}, which covers fundamental concepts related to Lie groups and Lie algebras.
\subsection{Lie Group and Lie Algebra}
A Lie group $\mathbb{G}$ is a mathematical structure that combines the properties of a group with those of a smooth manifold, either real or complex. In this structure, the operations of multiplication and inversion are smooth mappings on the manifold.

A Lie algebra $\mathfrak g$ is defined as the vector space comprised of the tangent vectors at the identity element of a Lie group, equipped with a bilinear operation known as the Lie bracket $[\cdot, \cdot] \in \mathfrak g$. For any elements in a Lie algebra need to satisfy the following operations:
\begin{enumerate}
    \item \textbf{Jacobi identity:} $\forall A, B, C \in \mathfrak g$,
    \begin{equation}
        [A, [B,C]] + [B,[C,A]] + [C,[A,B]] = 0
        \label{eq: jacobi}
    \end{equation}
    \item \textbf{Anticommutative:} $\forall A,B \in \mathfrak g$,
    \begin{equation}
        [A,B]=-[B,A]
        \label{eq: anticommut}
    \end{equation}
\end{enumerate}
For a matrix Lie group, the Lie bracket is considered as: 
\begin{equation}
    [A, B] = AB-BA
    \label{eq: liebracket}
\end{equation}
\subsection{$\mathbb{SU}(2)$ Group}
The $\mathbb{SU}(2)$ group consists of $2\times 2$ unitary matrices with determined one:
\begin{equation}
    \mathbb{SU}(2) = \left\{U \in \mathbb{C}^{2\times2} |U^{\dagger}U=I,\quad \textrm{det}(U)=1\right\}
    \label{eq: def_SU}
\end{equation}

The Lie algebra $\mathfrak{su}(2)$ is spanned by the three Pauli anti-Hermitian matrices $iX$, $iY$, and $iZ$, shown in Equation \eqref{eq: su_lie_algebra}, which form a basis for the tangent space of the $\mathbb{SU}(2)$ at the identity element. This means that any element $A$ in $\mathfrak{su}(2)$ space can be presented as a linear combination of the basis elements $iX$, $iY$, and $iZ$, shown in Equation \eqref{eq: ele_in_su}.
\begin{equation}
    \mathfrak{su}(2)=\textrm{span}\{iX,iY,iZ\}
    \label{eq: su_lie_algebra}
\end{equation}
\begin{equation}
    A= i(\alpha X+\beta Y+\gamma Z), \quad \alpha,\beta,\gamma\in R
    \label{eq: ele_in_su}
\end{equation}

Given the mapping that exists between Lie algebra and their corresponding Lie group, the element $A$ in $\mathfrak{su}(2)$ can be mapped to an element $\mathbb{G}_A$ of the Lie group $\mathbb{SU}(2)$, as shown in Equation~\eqref{eq: exp_map}.
\begin{equation}
\begin{aligned}
    \mathbb{G}_A &= \exp(A)=\exp\{i(\alpha X+\beta Y+\gamma Z)\}\\
    & = e^{i(\alpha X+\beta Y+\gamma Z)},\quad \alpha,\beta,\gamma\in R 
    \label{eq: exp_map}
\end{aligned}
\end{equation}
\subsection{Bi-invariant Riemannian metric}
The bi-invariant Riemannian metric refers to a Riemannian metric $Q\left \langle \cdot,\cdot \right \rangle$ on a Lie group $\mathbb{G}$ that is invariant under both left and right translation. For a given $g\in \mathbb{G}$, the left translation is defined as $L_{g}(h)=gh$ for any element $h \in \mathbb{G}$, and similarly, the right translation is defined as $R_{g}(h)=hg$. Furthermore, the differential forms of the left and right translations are denoted by $dL_{g}$ and $dR_{g}$, respectively. These maps represent transformations between tangent spaces at different points on the Lie group. Specifically, for a given $g\in \mathbb{G}$,
\begin{equation}
\begin{aligned}
    dL_{g}:T_{e}G \rightarrow T_{g}G \\
    dR_{g}:T_{e}G \rightarrow T_{g}G
    \label{eq: dif_invariant}
\end{aligned}
\end{equation}
Therefore, for the bi-invariant Riemannian metric, for any $A, B
\in \mathfrak{g}$,
\begin{equation}
    \langle dL_{g}(X), dL_{g}(Y)\rangle_{g} = \langle X,Y \rangle_{e} = \langle dR_{g}(X), dR_{g}(Y) \rangle_{g}
    \label{eq: bi-invariant}
\end{equation}

\textbf{definition} 
Considering a Lie algebra $\mathfrak{g}$ and Lie group $\mathbb{G}$, $Ad$ is the adjoint representation on Lie group, so we have $Ad_{g}(X)=gXg^{-1},g\in \mathbb{G},\quad X \in \mathfrak{g}$. For any element $X, Y \in \mathfrak{g}$ and $g\in\mathbb{G}$, if $\langle Ad_{g}X,Ad_{g}Y \rangle=\langle X,Y \rangle$, then it so called Ad-invariant inner product.

\textbf{theorem}
There exists a bi-invariant Riemannian metric on $\mathbb{SU}(2)$.

\begin{proof}{Proof}
The Killing form\footnote{Named after Wilhelm Killing} of a Lie algebra is defined as $B(X,Y)=tr(ad(X)\circ ad(Y))$

The Killing form of $\mathfrak{su}(2)$ is $B(X,Y)=-tr(XY)$. Because $\langle Ad_{g}X,Ad_{g}Y\rangle =-tr(gXg^{-1}gYg^{-1}) =-tr(gXYg^{-1}) =-tr(XY) =\langle X,Y\rangle $ therefore, the Ad-invariant inner product can be defined as $\langle X,Y\rangle=-tr(XY)$

Since the metric is defined by left-invariant translation, it naturally satisfies the left-invariance property. For the right-invariant translation, we have $\langle dR_{g}X, dR_{g}Y\rangle =\langle Xg,Yg\rangle_{g}=\langle g^{-1}Xg,g^{-1}Yg\rangle_{e} =-tr(g^{-1}XYg)=-tr(XY) =\langle X,Y\rangle$

\end{proof}

\subsection{Notation}
For convenience and reference, we collect all the main mathematical symbols used in this article in Table \ref{tab:notation}:

\begin{table}[H]
    \caption{Summary of notation}
    \label{tab:notation}
    \centering
        \begin{tabular}{cl}
    \hline
    Symbol & Meaning \\ 
    \hline
    $\mathbb{SU}(2)$ & Special unitary group with determinant 1 \\
    $\mathfrak{su}(2)$ & Lie algebra of traceless skew-Hermitian matrices \\
    $Ad$ & The adjoint representation on Lie group \\
    $ad$ & The adjoint operation on Lie algebra\\
    $I$ & Identity matrix \\
    $X, Y, Z$ & Pauli matrices \\
    $U_{0}$ & Any initialization element in $\mathbb{SU}(2)$ \\
    $U(t)$ & Geodesics from any starting element \\
    $L_{g},R_{g}$ & Left- and right-invariant operation \\
    $dL_{g},dR_{g}$ & Differential forms of the left- and right-invariant operation \\
    $T_eG,T_{g}G$ & Tangent space of element $e,g\in \mathbb{G}$ \\
    $\theta(t)$ & Time-dependent function of quantum gates parameters \\
    $v$ & First derivative of $\theta(t)$ (velocity) \\
    $\overline{v}$ & Average velocity \\
    $a$ & Second derivative of $\theta(t)$ (acceleration) \\
    $\overline{a}$ & Average acceleration \\
    $\tau$ & Relative standard deviation \\
    $E$ & Defined energy of the trajectory \\
    $L$ & Defined length of the trajectory \\
    \hline
\end{tabular}
\end{table}

\section{Design and Method}\label{sec: 3}
\begin{figure}[htbp]
    \centering
    \includegraphics[scale=0.32]{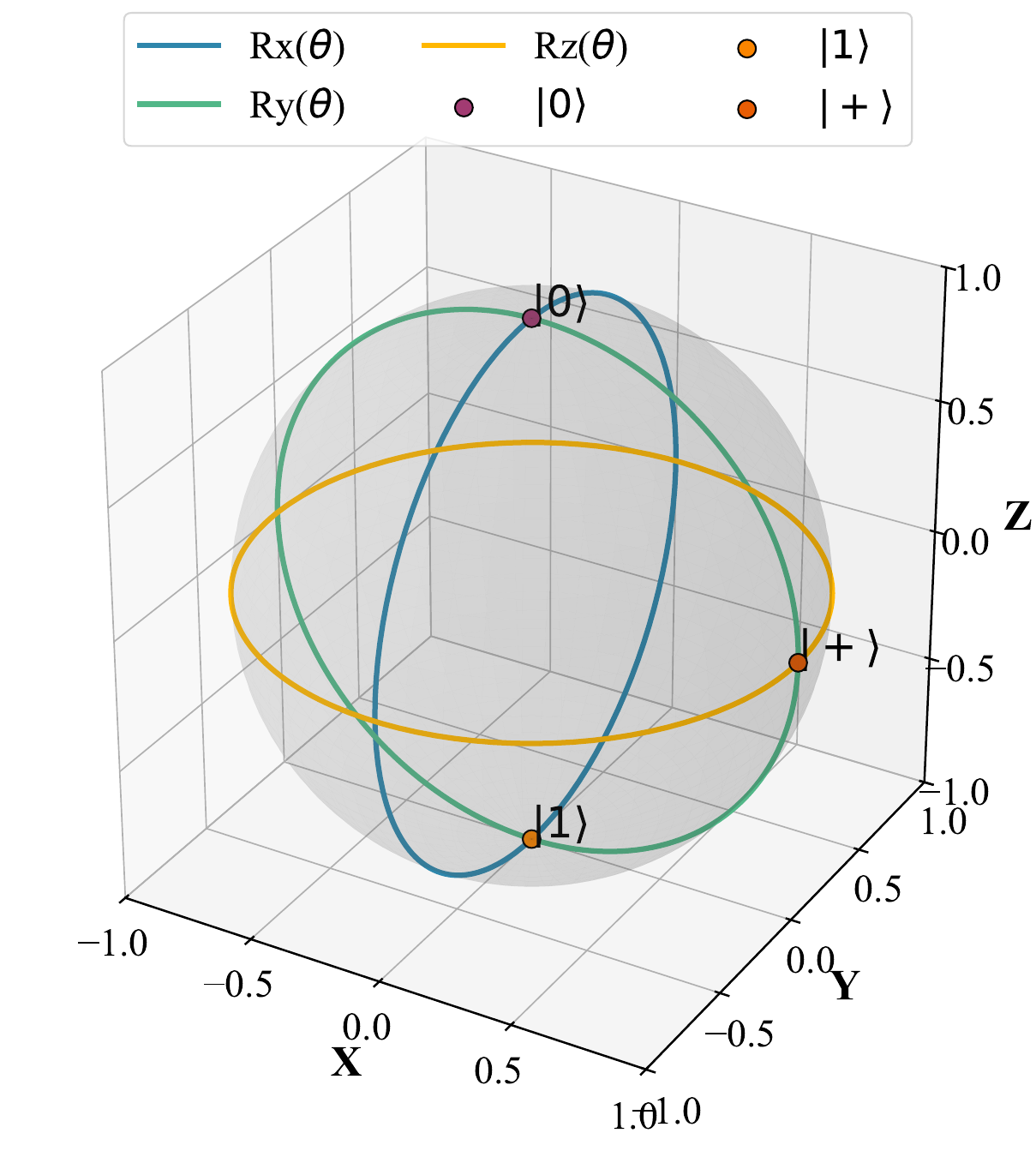}
    \caption{Geometric representation of single-qubit rotation gates on the Bloch sphere. These trajectory represents the operation of $R_{x}(\theta),R_{y}(\theta),R_{z}(\theta)$ as the element of $\mathbb{SU}(2)$. The $R_{x}$ and $R_{y}$ is act on the $\ket{0}$ whill the $R_{z}$ is act on the $\ket{+}=\frac{\ket{0}+\ket{1}}{\sqrt{2}}$.}
    \label{fig:bloch}
\end{figure}

The design and implementation of single-qubit quantum gates are deeply rooted in the mathematical structure of the special unitary group $\mathbb{SU}(2)$ and its associated Lie algebra $\mathfrak{su}(2)$. A key aspect of this relationship is the exponential mapping from $\mathfrak{su}(2)$ to $\mathbb{SU}(2)$, which provides a natural parametrization of all possible single-qubit operations. As shown in detail in Equation~\eqref{eq: rotation_gates},  the structure of commonly used single-qubit gates closely aligns with the general form of this mapping, differing only in the choice of coefficients within the linear combination of basis elements. This correspondence formally establishes the interpretation of single-qubit gates as elements of the $\mathbb{SU}(2)$ group. To provide a more intuitive illustration of the action trajectory of single-qubit quantum gates, we visualize their evolution on the Bloch sphere, as shown in Figure \ref{fig:bloch}.
\begin{equation}
\begin{aligned}
    R_x(\theta) & =\exp(-iX\theta/2), \quad \alpha=-\theta/2,\beta=\gamma=0 \\
    R_y(\theta)& =\exp(-iY\theta/2),\quad \beta=-\theta/2,\alpha=\gamma=0
    \\
    R_z(\theta) &=\exp(-iZ\theta/2), \quad \gamma=-\theta/2,\alpha=\beta=0
    \label{eq: rotation_gates}
\end{aligned}
\end{equation}

For geodesics on $\mathbb{SU}(2)$ group, we begin by considering the case of the identity element $I$. Under the exponential map, an element $A\in \mathfrak{su}(2)$, when evolved with time $t$, gives rises to a curve on $\mathbb{SU}(2)$. This curve, given by $\exp{(tA)}$, forms a one-parameter subgroup of the $\mathbb{SU}(2)$, as shown in Equation~\eqref{eq: exp_map_of_A}.
\begin{equation}
\begin{aligned}
    U(t) &=\exp{(tA)},A\in \mathfrak{su}(2) \\
    U(0) &=\exp{(0\cdot A)}=I
    \label{eq: exp_map_of_A}
\end{aligned}
\end{equation}

The one-parameter subgroup $\exp{(tA)}$ not only represents a continuous evolution within the Lie group $\mathbb{SU}(2)$, but also corresponds to a geodesic path under the standard bi-invariant Riemannian metric on the group \citep{18}. For any initial element $U_{0} \in \mathbb{SU}(2)$, the trajectory moving along with the direction $A \in \mathfrak{su}(2)$ can be defined via left translation as shown in Equation~\eqref{eq: left_translation_U}.

\begin{equation}
    \begin{aligned}
            &L_{\mathbb{SU}(2)}: \mathbb{SU}(2) \rightarrow \mathbb{SU}(2)
    \\
    &L_{\mathbb{SU}(2)}U(t) =U_{0}U(t)
    \label{eq: left_translation_U}
    \end{aligned}
\end{equation}

Due to the left-invariant vector, the characteristic of the geodesic remains unchanged. We can represented the geodesic along the $A$ direction as Equation~\eqref{eq: exp_tA}.
\begin{equation}
    U(t)=U_{0}\exp(tA)
    \label{eq: exp_tA}
\end{equation}

In neural network-generated quantum gate parameters for barren plateau mitigation methods, the parameters change with time. Hence, the parameter $\theta$ can be expressed as a time-dependent function, denoted as $\theta(t)$. A single-qubit gate can be considered as an element of $\mathbb{SU}(2)$, and it is constructed from the basis elements of its associated Lie algebra $\mathfrak{su}(2)$. Each quantum gate depends on a specific basis, indicating that a particular direction has been selected within the algebra. Therefore, the study of transformations of single-qubit gates in $\mathfrak{su}(2)$ can be reduced to analyzing the parameter $\theta$ transformations associated with these gates. We consider a trajectory generated by a time-dependent element of the Lie algebra:
\begin{equation}
    A(t)=\theta(t)A
    \label{eq: A_theta_A}
\end{equation}
where $A\in \mathfrak{su}(2)$ is the selected direction in Lie algebra(\emph{e.g.} $X, Y, Z$) and $\theta(t)$ is the parameters function of time that control the changes along with the direction. The corresponding trajectory can be expressed as:
\begin{equation}
    U(t)=U_{0}\exp \left( \int_{0}^{t}\theta(s)Ads \right)
    \label{eq: int_A}
\end{equation}

When $\theta(t)$ varies slowly, which means the variations in $\theta(t)$ remain close to zero. This trajectory is close to the standard geodesic, as shown in Equation~\eqref{eq: approx_geodesic}. 
\begin{equation}
    U(t)\approx U_{0}\exp(t\theta_{0}A), \theta_{0}=\theta(0)
    \label{eq: approx_geodesic}
\end{equation}
For this reason, the variation $\Delta \theta$ must be sufficiently small. This implies that the first derivative of $\theta(t)$ should be constant or approximately zero, indicating that the parameter remains nearly unchanged. Meanwhile, the second derivative should also be close to zero, which suggests that the trend is stable without significant curvature or acceleration. For clarity, we define the first derivative as the velocity $v$ and the second derivative as the acceleration $a$. Due to the discrete nature of the data, derivatives are approximated using central difference quotients derived from neighboring data points:
\begin{equation}
\begin{aligned}
    v &=\frac{d\theta(t)}{dt} = \frac{\theta(t+\delta{t})-\theta(t-\delta{t})}{2\delta{t}} \\
    a &=\frac{dv}{dt}=\frac{v(t+\delta{t})-v(t-\delta{t})}{2\delta{t}}
    \label{eq: velo_accler}
\end{aligned}
\end{equation}

We employ the relative standard deviation (RSD), defined as the ratio of the standard deviation $\sigma$ to the mean $\mu$: 
\begin{equation}
    \tau = \frac{\sigma}{\mu}
\end{equation}
In addition, we compute the optimization energy and trajectory length of the data: 
\begin{equation}
\begin{aligned}
        E &= \frac{\Sigma_{1}^{n}|\theta_{i}|^{2}}{n} \\
        L &= \Sigma_{1}^{n-1}\left|\theta_{i+1}-\theta_{i}\right|
\end{aligned}
\label{eq: E_and_L}
\end{equation}
The energy is given by the average of the squared values across all data points, while the path length is obtained by summing the absolute differences between each data point.
\begin{figure*}[t]
  \centering
  \subfloat[4 Qubits]{
    \includegraphics[width=0.48\textwidth]{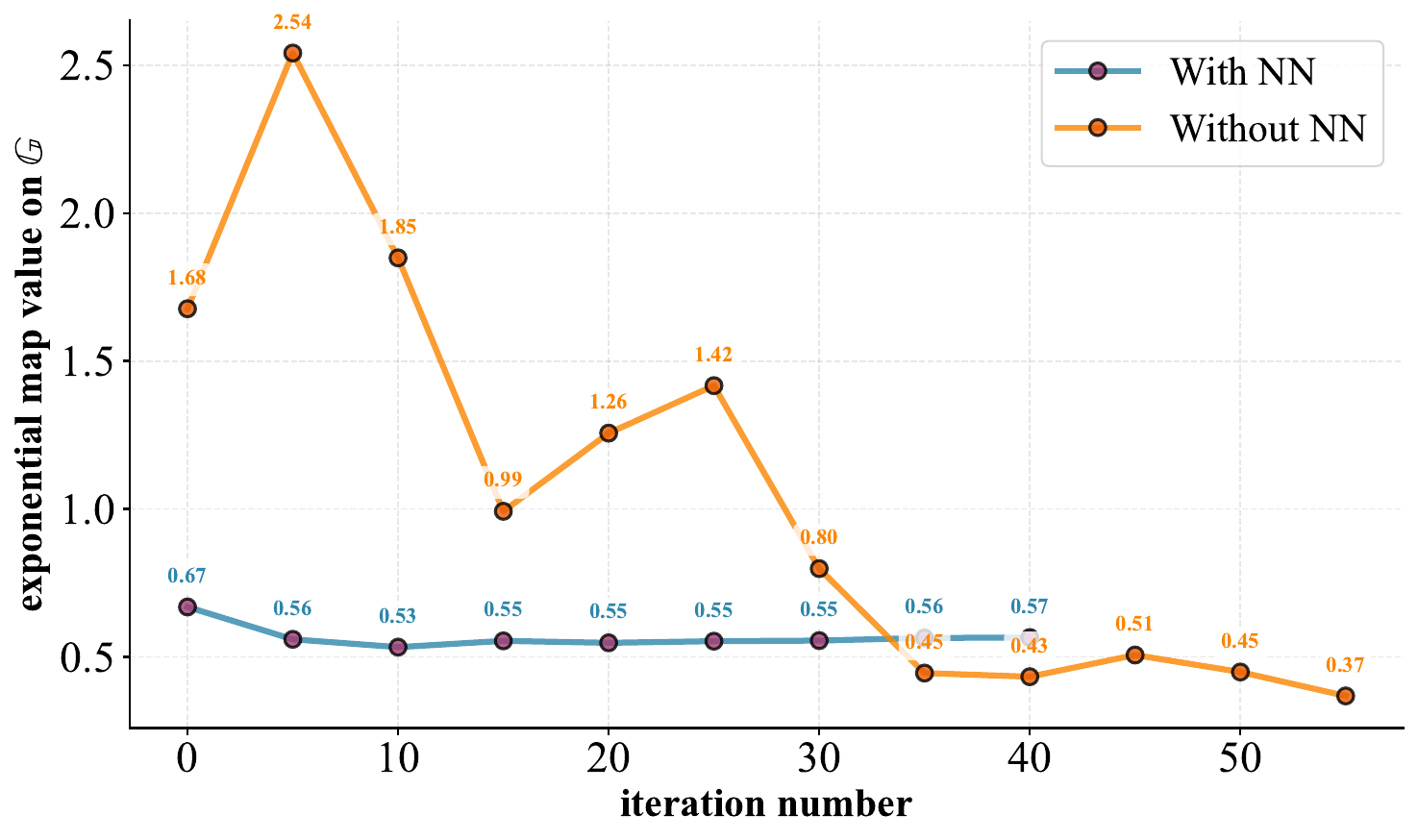}
    \label{fig:4q_map}
  }
  \hfill
  \subfloat[5 Qubits]{
    \includegraphics[width=0.48\textwidth]{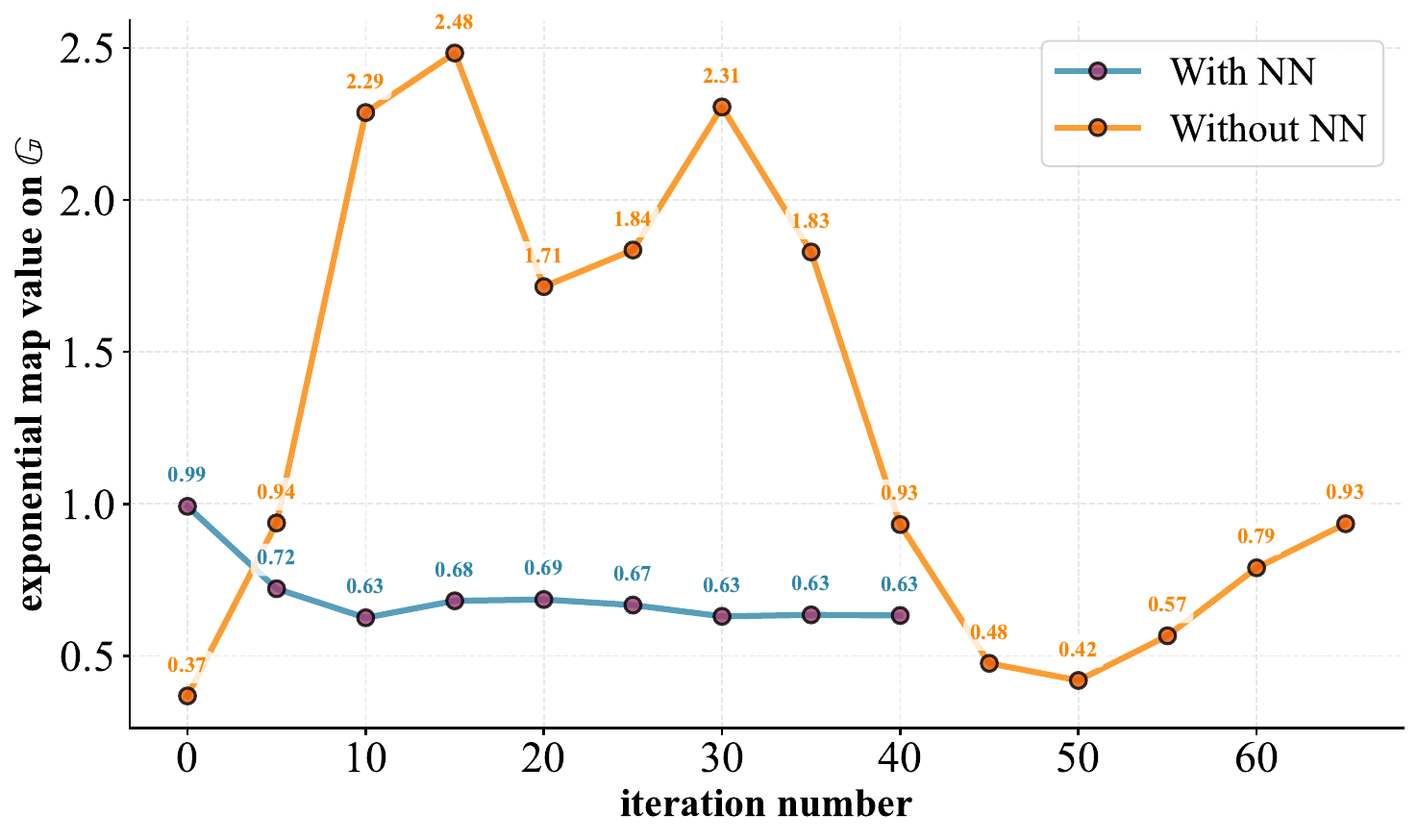}
    \label{fig:5q_map}
  }
  
  \vspace{1em}

  \subfloat[6 Qubits]{
    \includegraphics[width=0.48\textwidth]{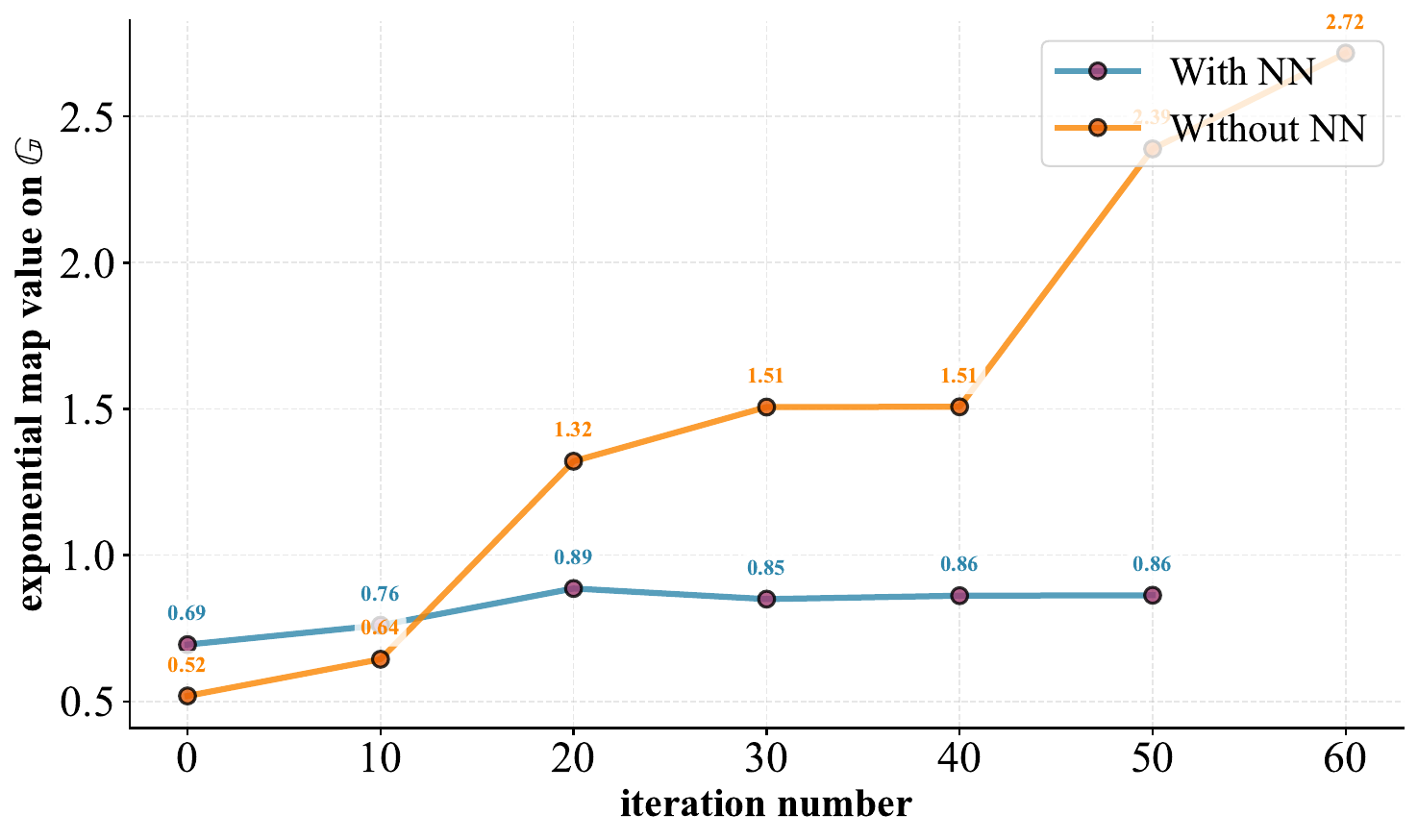}
    \label{fig:6q_map}
  }
  \hfill
  \subfloat[7 Qubits]{
    \includegraphics[width=0.48\textwidth]{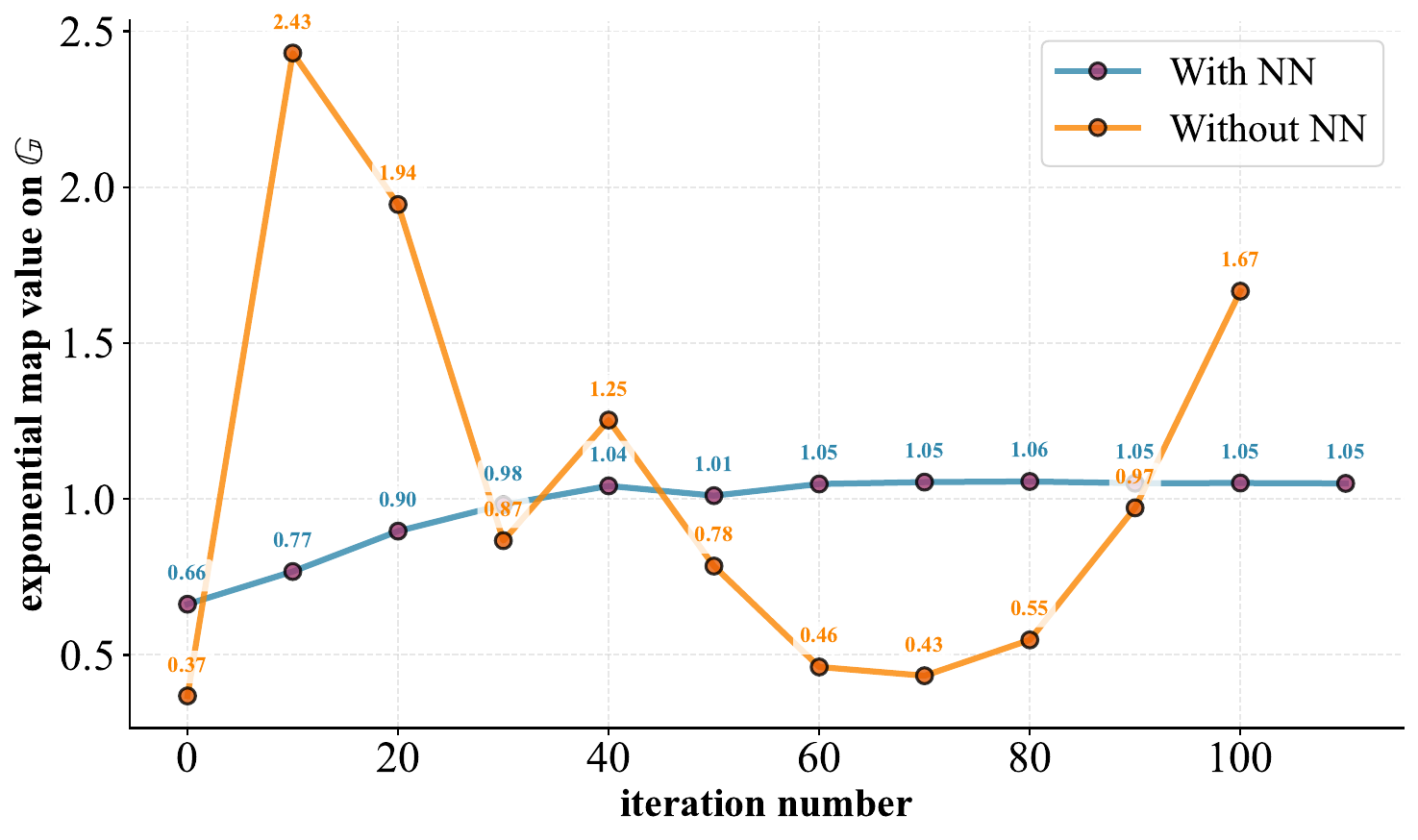}
    \label{fig:7q_map}
  }

  \vspace{1em}

  \subfloat[8 Qubits]{
    \includegraphics[width=0.48\textwidth]{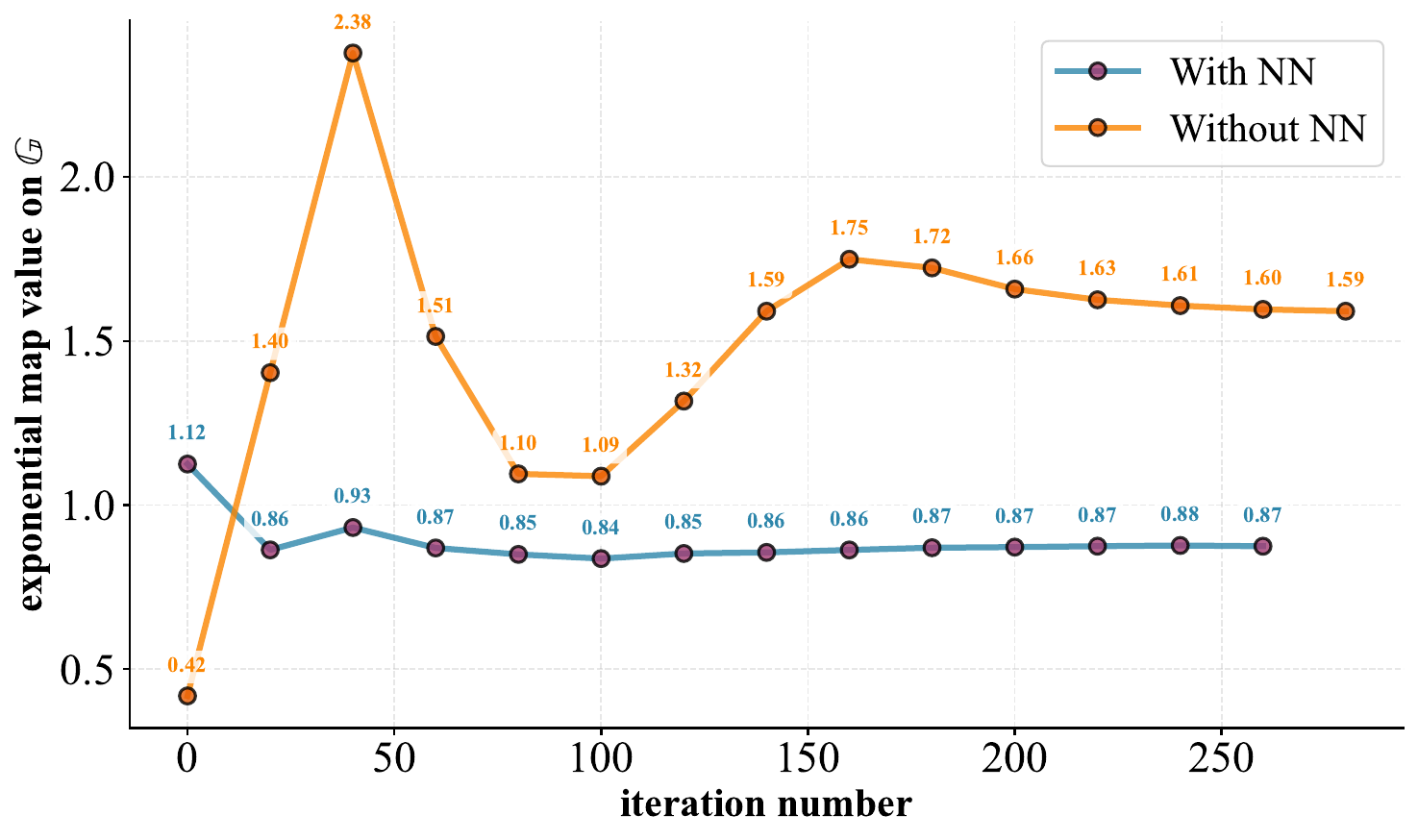}
    \label{fig:8q_map}
  }
  \hfill
  \subfloat[9 Qubits]{
    \includegraphics[width=0.48\textwidth]{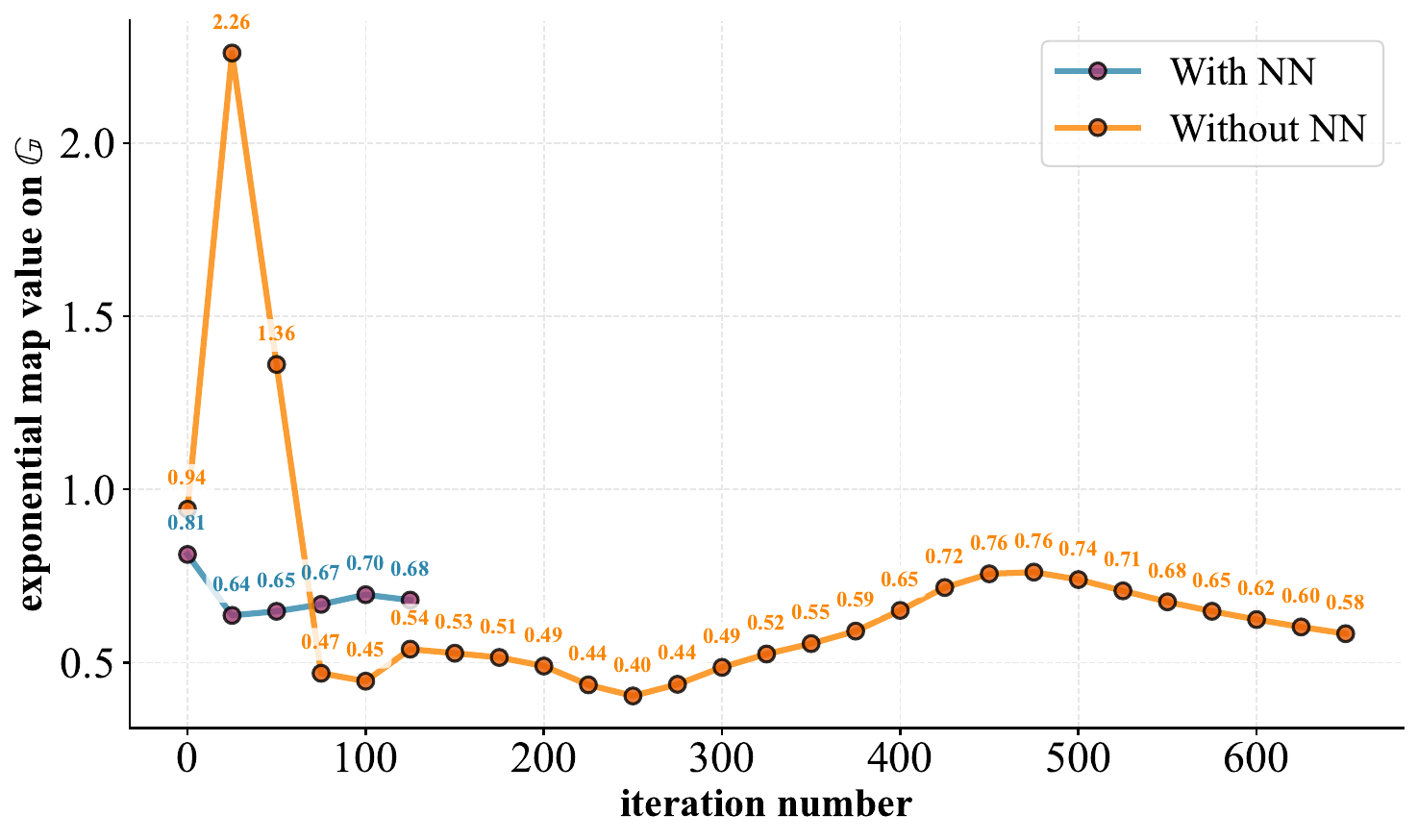}
    \label{fig:9q_map}
  }

  \caption{Fluctuation comparison of the group-mapped value with and without the neural-network module. 
  The x-axis represents the iteration number $k$, and the y-axis represents the group-mapped value $y$. 
  We focus on the fluctuation pattern of $y$ across $k$. 
  Using neural networks helps mitigate barren plateaus \citep{15, 16}, resulting in fewer iterations to convergence and a shorter trajectory.}
  \label{fig:map_line}
\end{figure*}

\begin{figure*}[t]
  \centering
  \subfloat[4 Qubits]{
    \includegraphics[width=0.48\textwidth]{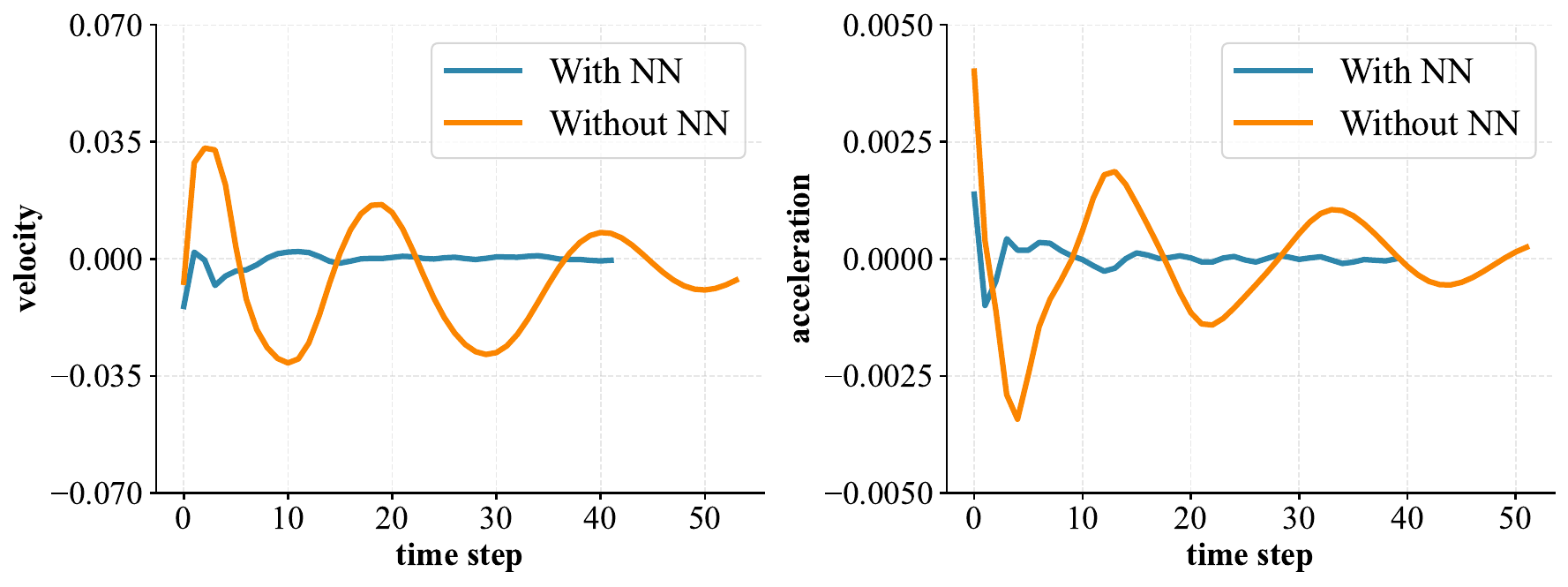}
  }
  \hfill
  \subfloat[5 Qubits]{
    \includegraphics[width=0.48\textwidth]{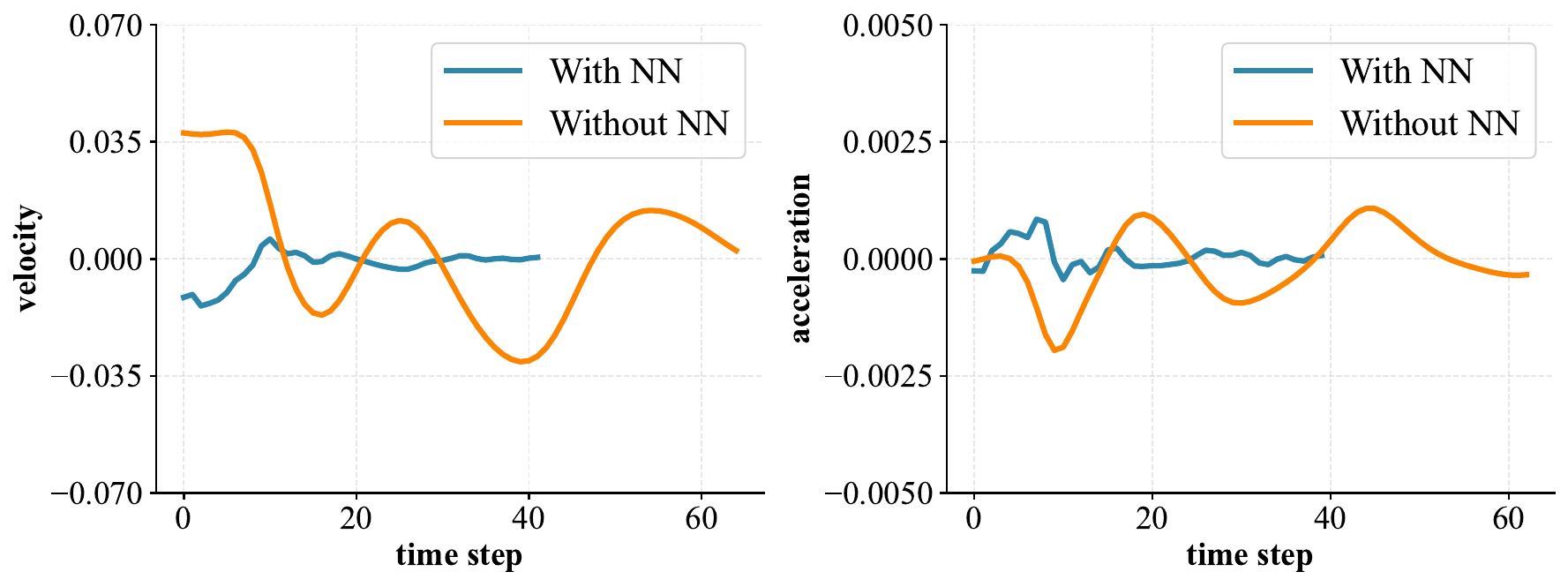}
  }
  
  \vspace{1em}

  \subfloat[6 Qubits]{
    \includegraphics[width=0.48\textwidth]{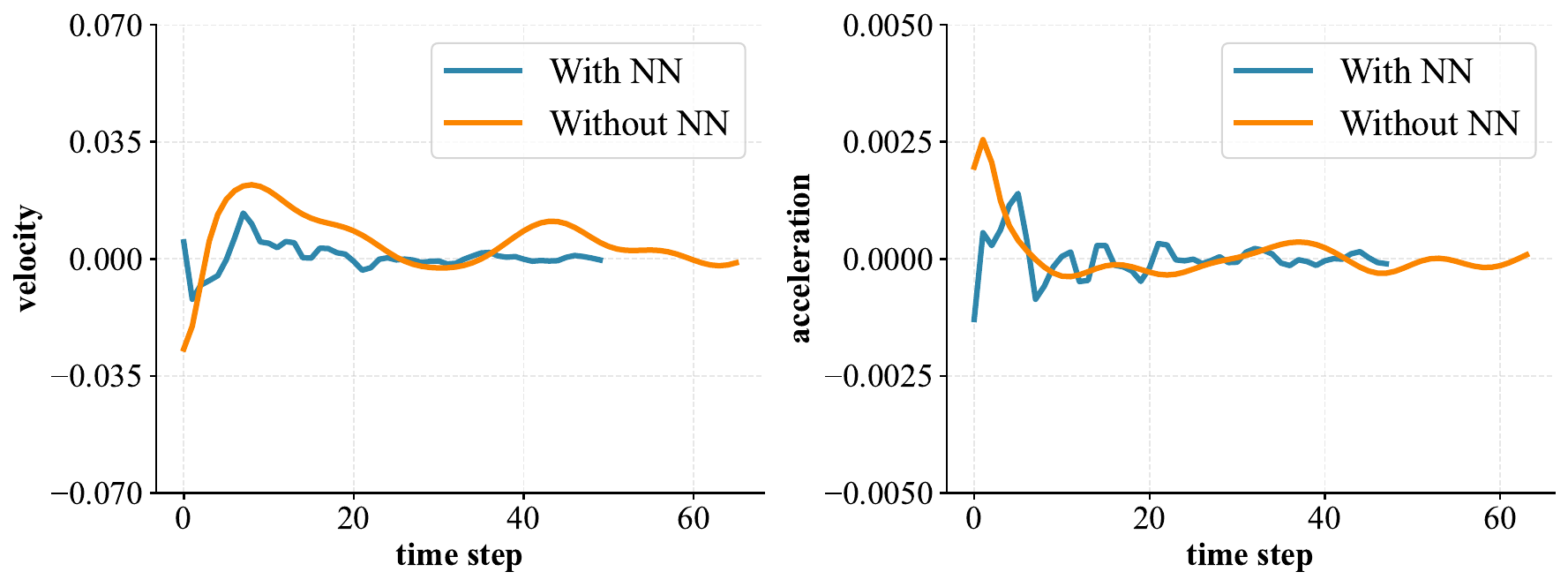}
  }
  \hfill
  \subfloat[7 Qubits]{
    \includegraphics[width=0.48\textwidth]{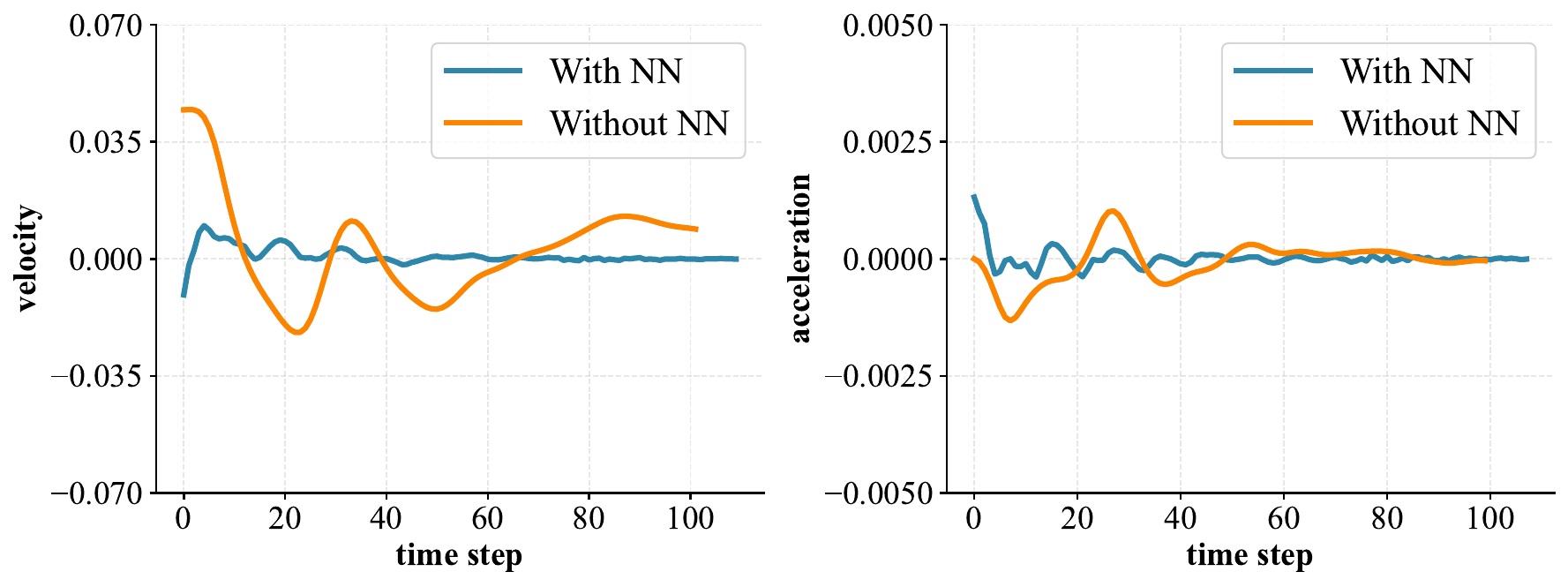}
  }
  
  \vspace{1em}

  \subfloat[8 Qubits]{
    \includegraphics[width=0.48\textwidth]{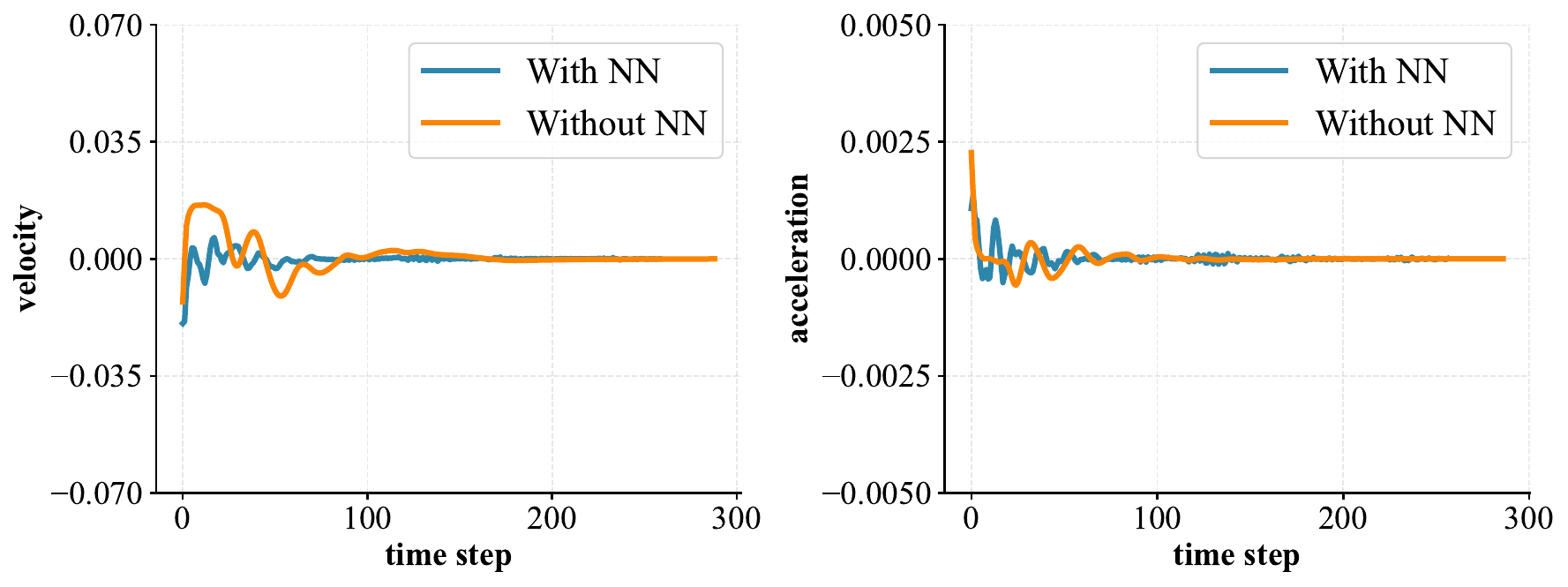}
  }
  \hfill
  \subfloat[9 Qubits]{
    \includegraphics[width=0.48\textwidth]{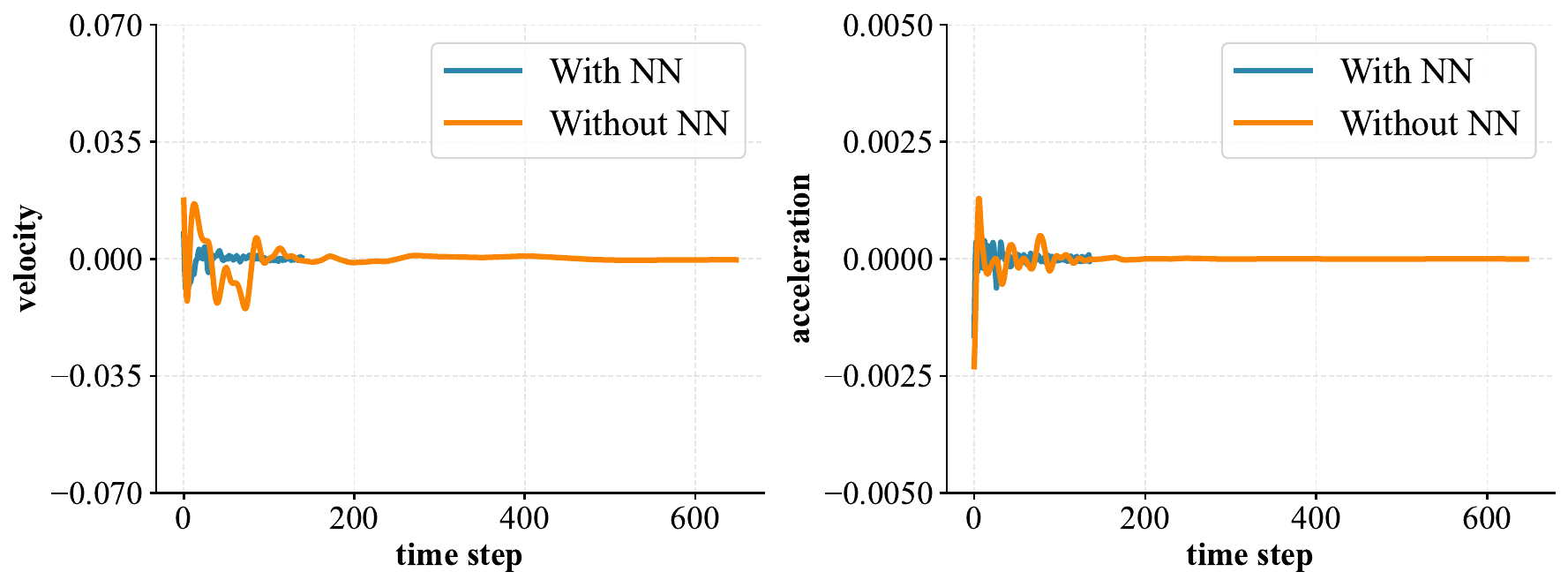}
  }

  \caption{Comparison of the Velocity and Acceleration in Generating Parameters with Different Numbers of Qubits. 
  The neural network-enhanced model (left/blue usually) shows velocity and acceleration closer to zero, indicating geodesic-like stability.}
  \label{fig:va_line}
\end{figure*}

\begin{table*}[htbp]
    \centering
    \caption{Comparison of Parameters Generated With and Without Using a Neural Network}
    \label{tab:parameters_comparison}
    \scalebox{1.0}{
    \begin{tabular}{lccccccc}
        \toprule
        \textbf{Qubits} & \textbf{Label} & \textbf{$\overline{v}$} & \textbf{$\overline{a}$} & \textbf{$E$} & \textbf{$L$} & \textbf{$\tau_{V}$} & \textbf{$\tau_{A}$} \\
        \midrule
        \multirow{2}{*}{4} & With NN & 1.4$\times10^{-3}$ & 1.6$\times10^{-4}$ & 8.0$\times10^{-6}$ & 4.3$\times10^{-1}$ & 2.01\% & 1.96\% \\
         ~ & Without NN & 1.4$\times10^{-2}$ & 8.9$\times10^{-4}$ & 2.97$\times10^{-4}$ & 4.12 & 1.16\% & 1.36\% \\
        \midrule
        \multirow{2}{*}{5} & With NN & 3.1$\times10^{-3}$ & 1.98$\times10^{-4}$ & 2.5$\times10^{-5}$ & 6.99$\times10^{-1}$ & 1.51\% & 1.38\% \\
         ~ & Without NN & 1.6$\times10^{-2}$ & 5.7$\times10^{-4}$ & 3.9$\times10^{-4}$ & 5.396 & 1.20\% & 1.26\% \\
        \midrule
        \multirow{2}{*}{6} & With NN & 2.5$\times10^{-3}$ & 2.7$\times10^{-4}$ & 1.6$\times10^{-5}$ & 0.81$\times10^{-1}$ & 1.58\% & 1.56\% \\
         ~ & Without NN & 7.6$\times10^{-3}$ & 3.1$\times10^{-4}$ & 1.03$\times10^{-4}$ & 2.57 & 1.16\% & 1.79\% \\
        \midrule
        \multirow{2}{*}{7} & With NN & 1.4$\times10^{-3}$ & 1.03$\times10^{-4}$ & 7$\times10^{-6}$ & 8.7$\times10^{-1}$ & 1.78\% & 2.04\% \\
        ~ & Without NN & 1.2$\times10^{-2}$ & 3.2$\times10^{-4}$ & 2.4$\times10^{-4}$ & 6.096 & 1.29\% & 1.39\% \\
        \midrule
        \multirow{2}{*}{8}& With NN & 8.1$\times10^{-4}$ & 7.7$\times10^{-5}$ & 5$\times10^{-6}$ & 1.27 & 2.71\% & 2.35\% \\
        ~ & Without NN & 2.6$\times10^{-3}$ & 6.6$\times10^{-5}$ & 2.4$\times10^{-5}$ & 3.79 & 1.88\% & 2.95\% \\
        \midrule
        \multirow{2}{*}{9} & With NN & 1.2$\times10^{-3}$ & 9.7$\times10^{-5}$ & 5$\times10^{-6}$ & 1.07 & 1.83\% & 1.99\% \\
        ~ & Without NN & 1.6$\times10^{-3}$ & 4.6$\times10^{-5}$ & 1.2$\times10^{-5}$ & 5.29 & 2.15\% & 3.69\% \\
        \bottomrule
    \end{tabular}
    }
\end{table*}
\section{Result}
\label{sec: 4}
The data used in this study were generated by running code developed in a previous work as shown in (see \citep{16}). In these experiments, we considered systems with 4 to 9 qubits, performing ten independent runs for each qubit number and recording the updated parameters in each case. We set $\delta t=5$, and to ensure consistency with the parameter range generated by the neural network, we normalized all non-neural network-generated data to the interval [$-1$, $1$] before experimentation. As results across the ten runs were closely consistent, we selected a representative result for each qubit number.

Figure \ref{fig:map_line} visualizes the resulting parameters as a curve. We sampled data points using different numbers of qubits and varying step lengths to avoid visual mass caused by excessive data density. Step lengths were set to 25 for 9 qubits, 20 for 8 qubits, 10 for 6-7 qubits, and 5 for 4–5 qubit systems. After applying an exponential transformation, the parameters generated by the neural network exhibit behavior more closely resembling a smooth curve, with minimal fluctuation. In contrast, the parameters obtained without the use of a neural network display significantly more irregularity. This characteristic explains smoother parameter updating when neural networks are used. From this perspective, updating the reflection parameters using a neural network aligns with the structure of the geodesic.

Figure \ref{fig:va_line} presents the velocity and acceleration of the parameter trajectories. We compare these quantities with those obtained from a model without the neural network. It is illustrated that, in the neural network-enhanced model, the velocity remains approximately zero, indicating that the slope of the geodesic is constant and shows no significant variation. Furthermore, the acceleration is also close to zero, implying that the trend of parameter evolution is stable, with no noticeable curvature or acceleration on the $\mathfrak{su}(2)$. In contrast, the performance of the model without the neural network is significantly worse, particularly in terms of the velocity, which exhibits higher instability. This suggests that neural network-based models have greater consistency with the properties of geodesics during parameter updates, as compared to models without neural networks.

Table \ref{tab:parameters_comparison} provides a detailed numerical analysis of the parameters calculated by models with and without neural networks. The metrics include average velocity ($\overline{v}$), average acceleration ($\overline{a}$), optimization energy ($E$), trajectory length ($L$), and the RSD values of velocity ($\tau_{V}$) and acceleration ($\tau_{A}$). From the data presented in the table, the $\overline{a}$ of the model with NN are smaller, indicating a smoother, more stable update trajectory compared with the model without NN. Specifically, the model incorporating a neural network exhibits superior performance, characterized by lower optimization energy consumption and a shorter trajectory length compared to the model without a neural network, while the RSD values for both models are similar. But this reduction in $E$ and $L$ indicates that the NN model's updated trajectory of neural network-generated parameters is more closely aligned with the properties of geodesics on $\mathbb{SU}(2)$. Lower optimization energy and trajectory length also imply reduced curvature and shorter optimization paths, consistent with convergence acceleration in manifold optimization. This indicates that neural network generated parameters evolve along geodesics on $\mathbb{SU}(2)$, showing a manifold constrained optimization that helps mitigate barren plateaus.

\clearpage
\section{Conclusion}
\label{sec: 5}
In this paper, we establish the mathematical foundation for mitigating the barren plateau problem in variational quantum algorithms by employing neural networks to generate gate parameters, using tools from Lie theory. By mapping single-qubit gates to the $\mathbb{SU}(2)$ group and its Lie algebra $\mathfrak{su}(2)$, we show that parameter optimization corresponds to transformations on $\mathfrak{su}(2)$. Numerical analysis reveals that the evolution of neural network-generated parameters follows a trajectory consistent with a geodesic on $\mathbb{SU}(2)$, exhibiting low velocity and acceleration, and thus stable and nearly linear dynamics. In contrast, non-neural network parameter updates display significant fluctuations and higher energy and path length, despite comparable relative standard deviation. These results indicate that if parameter update methods are aligned with the geodesic, they offer a promising strategy for improving training efficiency in quantum machine learning. This geometric explanation may guide new parameter initialization or architecture design in variational quantum algorithms. While this work only employs the $\mathbb{SU}(2)$ group, it has not been extended to higher-dimensional representations to experimentally implement multi-qubit gate operations. This remains a direction for future research.


%
%
%


\section*{Acknowledgments}{This work was supported in part by a Graduate Research Assistantship at the AARC Lab, Department of Electrical and Computer Engineering, University of Alabama in Huntsville. The first author also wishes to thank Prof. Yanying Liang and Prof. Haozhen Situ for their valuable guidance and support during his earlier research project \citep{16}.
}



\bibliography{sample} 





\end{document}